\documentclass[11pt, preprint]{article}
\usepackage{color} \usepackage{cancel}
\usepackage{float} \usepackage{graphicx}
\usepackage{subfigure} \usepackage{caption}			
\usepackage{dcolumn}
\usepackage[toc,page]{appendix} \usepackage{authblk}	 		
\usepackage{indentfirst}	
\usepackage{authblk}	 		
\usepackage{multirow}     
\usepackage{hyperref}     
\usepackage{amsfonts, amssymb, amsmath}
\usepackage[left=2.5cm,right=2.5cm,top=2.9cm,bottom=2.9cm,a4paper]{geometry}
\usepackage[noadjust]{cite}
\usepackage{filecontents}

\graphicspath{{./figures/}}
\hypersetup{ colorlinks   = true, 
urlcolor     = blue, 
linkcolor    = blue, 
citecolor    = red 	 
}
\newcommand{\bea}{\begin{eqnarray}} \newcommand{\ena}{\end{eqnarray}}
\newcommand{\beq}{\begin{equation}} \newcommand{\eeq}{\end{equation}}
           \newcommand{\nn}{\nonumber \\}

           \renewcommand{\l}{\lambda}
                
           \renewcommand{\O}{\Omega}

            \renewcommand{\t}{\tau}

\title{\Large\bf Primordial gravitational waves from the space-condensate inflation model}
\author[1]{Seoktae Koh\thanks{kundol.koh@jejunu.ac.kr}}
\author[2,3,4]{Bum-Hoon Lee\thanks{bhl@sogang.ac.kr}}
\author[4]{Gansukh Tumurtushaa\thanks{gansuh@sogang.ac.kr}}
\affil[1]{\small \it Department of Science Education, Jeju National University, Jeju, 63243, Korea }
\affil[2]{\small \it Asian Pacific Center for Theoretical Physics, Pohang 37673, Korea}
\affil[3]{\it Department of Physics, Postech, Pohang 790-784, Korea}
\affil[4]{\small \it Center for Quantum Spacetime \& Department of Physics, Sogang University, Seoul 121-742, Korea}

\date{\small }
\begin{document}
\maketitle


\begin{abstract}
We consider the space-condensate inflation model to study the primordial gravitational waves generated in the early Universe. We calculate the energy spectrum of gravitational waves induced by the space-condensate inflation model for the full frequency range with the assumption that the phase transition between two consecutive regimes is abrupt during the evolution of the Universe. The suppression of the energy spectrum is found in our model for the decreasing frequency of gravitational waves depending on the model parameter. To realize the suppression of the energy spectrum of the primordial gravitational waves, we study the existence of the early phase transition during inflation for the space-condensate inflation model.
\end{abstract}

\section{Introduction}
Inflation \cite{PhysRevD.23.347, PhysRevLett.48.1220, LINDE1982389} has been a very successful model because of its beautiful idea of explaining the early evolution of the Universe as well as its elegant way of solving the problems in standard big bang cosmology including the flatness problem, the horizon problem and the monopole problems. The basic idea of inflation is that the Universe went through a phase of accelerated expansion during its very early stage of evolution. As a result of this accelerated expansion, the inflation models predict that the seed for the large-scale structure and  the primordial gravitational waves can be generated \cite{Starobinsky:1979ty, PhysRevD.37.2078, PhysRevD.42.453}. Although the existence of the stochastic background of the primordial gravitational waves from inflation has not been observed yet, its observational detection would not only verify a success of inflation, but would also open a new window to understand the physics of the early Universe.

There are several observational constraints on the energy spectrum of the stochastic gravitational wave background at different frequency ranges; in the high-frequency range, with $f\in (10^{-4}, 10^4)$Hz, the gravitational waves can be detected by the space-based interferometric detectors such as eLISA/NGO \cite{elisa}, BBO \cite{BBO} and DECIGO \cite{0264-9381-28-9-094011} as well as the ground-based interferometric detectors such as Advanced LIGO \cite{0264-9381-27-8-084006}, VIGRO \cite{Accadia:2012zzb}, KAGRA \cite{0264-9381-27-8-084004}, and ET \cite{ET}; in the median-frequency range, with $f\in (10^{-9}, 10^{-7})$Hz, the gravitational waves can be probed by the pulsar timing experiments such as PTA \cite{2900303}, EPTA \cite{0264-9381-27-8-084014} and SKA \cite{SKA}; and for the low-frequency range, with $f<10^{-15}$Hz, the primordial gravitational waves can be detected by their unique imprint on the cosmic microwave background (CMB) power spectrum of the primordial B-mode polarization \cite{Kamionkowski:1996ks}. Detection of the primordial gravitational waves is the main goal of the CMB polarization experiments, such as EBEX \cite{ReichbornKjennerud:2010ja}, the SPT polarization \cite{Hanson:2013hsb}, and the ACT polarization \cite{Naess:2014wtr}.

In this work, in comparison to the single-field inflation model, we aim to calculate the energy spectrum of primordial gravitational waves generated by the space-condensate inflation model \cite{PhysRevD.88.043523, underprep}. The space-condensate inflation model is motivated by the nonlinear sigma models that came from the higher-dimensional gravity theories \cite{Omero:1980vx, GellMann:1984mu}. In the nonlinear sigma models, the sigma fields with $SO(3)$ symmetry are found to have solutions which are linearly proportional to the spatial coordinates, $\sigma^a\sim x^a$ $(a=1,2,3)$ \cite{ArmendarizPicon:2007nr, Endlich:2012pz, 'tHooft:2007bf, Lee:2009zv}. After taking the spatially linear configuration of $\sigma$ fields into account along with the Friedmann Lema$\hat{i}$tre-Robertson-Walker (FLRW) metric, the cosmological principles of homogeneity and isotropy are preserved for the resulting background evolution, and the space-condensate inflation model can be constructed.

In order to calculate the energy spectrum, for simplicity, we assume the abrupt phase transition between two consecutive regimes during evolution of the Universe. In such a case, the cosmic expansion is the only damping effect for the modes that reentered the horizon at late time, either the matter- or the radiation-dominated era. The outline of our paper is the following. After a brief overview of the formalism to obtain the energy spectrum of primordial gravitational waves in Sec.~\ref{sec2}, we calculate the energy spectrum of primordial gravitational waves for the space-condensate inflation model in Sect.~\ref{sec3}. Finally, we summarize our results in Sec.~\ref{sec4}.

\section{Review: Spectrum of the primordial gravitational waves }\label{sec2}
We begin our study with a brief review on the formalism to calculate the energy spectrum of the primordial gravitational waves that were produced during inflation \cite{Chongchitnan:2006pe, Zhao:2006eb, PhysRevD.77.063504, Liu:2015psa}. The gravitational waves are described by the tensor perturbation, $h_{ij}$, of the flat FLRW metric, and the perturbed metric is
\bea\label{FRWmetric}
ds^2=a^2(\tau)\left[-d\tau^2+(\delta_{ij}+h_{ij})dx^idx^j\right],
\ena
where $h_{ij}$ is symmetric under the exchange of indices, and satisfies the transverse-traceless condition.\,\footnote{$\partial_ih^{ij}=0$ and $\delta^{ij}h_{ij}=0$.} The cosmic time $t$ relates to the conformal time $\tau$ via a scale factor $a(\tau)$, and it is $dt= a(\tau)d\tau$.

Since the gravitational waves are very weak such that $|h_{ij}|\ll 1$, we study the linearized Einstein equation
\bea\label{GWevolEq}
-\frac{1}{2}h_{ij;\nu}\,^{;\nu}=\frac{1}{M_p^2}\Pi_{ij},
\ena
where $\Pi_{ij}$ is the anisotropic part of the stress tensor, and satisfies $\Pi^i\,_{i}=0$ and $M_p^2\equiv1/8\pi G$ is the reduced Planck mass.  The tensor perturbation and the anisotropic part of the stress tensor are expanded in the Fourier space as
\bea
h_{ij}(\tau,\mathbf{x})&=& \frac{\sqrt{2}}{M_p} \sum_{\lambda}\int\frac{d\,\mathbf{k}}{(2\pi)^{3/2}}\epsilon^{(\lambda)}_{ij}(\mathbf{k})h_{\mathbf{k}}^\lambda e^{i\mathbf{k}\mathbf{x}},\\
\Pi_{ij}(\tau,\mathbf{x})&=& \frac{\sqrt{2}}{M_p} \sum_{\lambda}\int\frac{d\,\mathbf{k}}{(2\pi)^{3/2}}\epsilon^{(\lambda)}_{ij}(\mathbf{k}) \Pi_{\mathbf{k}}^\lambda e^{i\mathbf{k}\mathbf{x}},
\ena
where the superscript $\lambda$ denotes each polarization state of the tensor perturbations and $\epsilon_{ij}^{(\lambda)}$ is the polarization tensor and satisfies $\epsilon_{ij}^{(\lambda)}\epsilon^{ij(\lambda')}=\delta^{\l\l '}$. Here, we denote $h_{\mathbf{k}}^{(\l)}(\t)$ and $\Pi_{\mathbf{k}}^{(\l)}(\t)$ as $h_{\l,k}(\t)$ and $\Pi_{\l, k}(\t)$, respectively. The linearized evolution equation of the gravitational waves described by Eq. (\ref{GWevolEq}) is written in the Fourier space as
\bea\label{waveeq}
h_{\l,k}''+2\mathcal{H}h_{\l,k}'+k^2h_{\l,k}=\frac{2}{M_p^2} a^2(\t)\Pi_{\l,k}\,,
\ena
where $\mathcal{H}=a'/a$ and $'\equiv d/d\t$. For simplicity, in this work, we ignore the anisotropic part of a stress tensor that generates the nonzero source term on the right-hand side of Eq. (\ref{waveeq}) in general.

One of the important predictions of inflation is that the primordial power spectrum for the scalar perturbation is almost Gaussian and nearly scale invariant. The power spectrum for the scalar perturbation can be parametrized as
\bea
\mathcal{P}_S(k)=\mathcal{P}_{S}(k_{\ast})\left(\frac{k}{k_{\ast}}\right)^{n_S-1+\frac{\alpha_S}{2}\ln (k/k_{\ast})},
\ena
where $\alpha_S\equiv dn_{S}/d\ln k$ is the running of the scalar spectral index $n_{S}$ and $k_{\ast}$ is the reference wave number.
Another key prediction of inflation is the existence of the primordial gravitational waves. The power spectrum of the primordial gravitational waves observed today, $P_{T}(k)$, is defined in the Fourier space as follows:
\bea\label{presentPS}
P_{T}(k)\equiv\frac{4k^3}{\pi^2 M_p^2}\sum_{\l}\langle h_{\l,\,k}^\dagger h_{\l,\,k}\rangle \,,
\ena
where the bracket, $\langle\ldots\rangle$, indicates the spatial average.

The strength of the primordial gravitational waves is characterized by their energy spectrum
\bea\label{gwEs}
\O_{GW}(k)=\frac{1}{\rho_{\text{crit}}}\frac{d\rho_{GW}}{d\ln k},
\ena
where $\rho_{\text{crit}}=3H_0^2M_p^2$ is the critical density and $H_0$ is the present Hubble constant. The energy density of the gravitational wave background, $\rho_{GW}$, is defined by $\rho_{GW}=-T^0\,_0$, and it is related to Eq. (\ref{presentPS}) as
\bea\label{rhoGW}
\rho_{GW}=\frac{M_p^2}{4}\int k^2 P_{T}(k) d\ln k.
\ena
By using Eq. (\ref{rhoGW}) and  Eq. (\ref{gwEs}), we obtain
\bea\label{gwEsofk1}
\O_{GW}(k)=\frac{k^2}{12H_0^2}P_{T}(k),
\ena
and it is consistent with Refs.~\cite{Chongchitnan:2006pe, Zhao:2006eb, PhysRevD.77.063504, Liu:2015psa}. The tensor power spectrum observed today can be related to the that of the inflationary one by the transfer function $\mathcal{T}(k)$ as follows
\bea
P_{T}=\mathcal{T}^2(k)\mathcal{P}_{T}(k),
\ena
where the transfer function $\mathcal{T}$ reflects the damping effect after modes reentered the horizon. By using this relation, we can reexpress Eq. (\ref{gwEsofk1}) in terms of the inflationary tensor power spectrum
\bea\label{gwEsofk}
h_0^2\O_{GW}(k)=\frac{h_0^2k^2}{12H_0^2}\mathcal{T}^2(k)\mathcal{P}_{T}(k),
\ena
where $h_0^2\O_{GW}(k)$ is the physically meaningful quantity for the gravitational waves. Thus, we shall calculate constraints on this quantity. In the remaining part of this section, we discuss $\mathcal{T}(k)$ and $\mathcal{P}_{T}(k)$.

As we mentioned earlier, the transfer function reflects the damping effect after modes reentered the horizon. Several damping effects have been studied previously in Refs. \cite{Zhao:2006eb, PhysRevD.77.063504, Liu:2015psa}. In this work, we consider the damping effect only due to the cosmic expansion since it shows the major contribution in the evolution of the Universe. Thus, the evolution of the primordial gravitational waves can be described by the following equation:
\bea\label{waveeq1}
h_{\l,k}''+2\mathcal{H}h_{\l,k}'+k^2h_{\l,k}=0,
\ena
and it only depends on the evolution of the scale factor. The mode solutions to this equation have qualitative behavior in two regimes as it evolves in time: far outside the horizon $(k\ll aH)$ where the amplitude of $h_{\l,k}$ stays constant, and far inside the horizon $(k\gg aH)$ where they damp as $h_{\l, k}\sim 1/a$.

The transfer function for the modes which are well inside the horizon ($k\gg10^{-18}$ Hz) has been calculated by integrating Eq. (\ref{waveeq1}) numerically from $\tau=0$ to $\tau_0$; a good fit to the transfer function due to Ref.~\cite{Turner:1993vb} is
\bea\label{waveeq2}
\mathcal{T}(k)&=&
\frac{3}{k^2\tau_0^2}\frac{\O_m}{\O_{\Lambda}}\sqrt{1+\frac{4}{3}\left(\frac{k}{k_{eq}}\right)+\frac{5}{2}\left(\frac{k}{k_{eq}}\right)^2},
\ena
where $k_{eq}$ is the wave number corresponding to a Hubble radius at the time that matter and radiation have equal energy density, $\tau_0$ is the present conformal time, and $\O_m$ and $\O_\Lambda$ are the present energy density fractions of the matter and the vacuum, respectively. The factor $\O_m/\O_\Lambda$ represents the effect of the accelerated expansion of the Universe \cite{PhysRevD.74.043503}. For modes that reentered the horizon during the matter-dominated era in which $k\ll k_{eq}$, Eq. (\ref{waveeq2}) evolves as $\mathcal{T}(k)\sim k^{-2}$, while it evolves as $\mathcal{T}(k)\sim k^{-1}$ for modes that reentered the horizon during the radiation-dominated era in which $k\gg k_{eq}$.

By combining Eq. (\ref{gwEsofk}) with Eq. (\ref{waveeq2}), we obtain the energy spectrum of the gravitational waves as
\bea\label{h02OGW}
h_0^2\O_{GW}(k)\simeq\frac{3h_0^2}{4H_0^2\tau_0^4}\left(\frac{\O_m}{\O_{\Lambda}}\right)^2 \frac{1}{k^2} \left[1+\frac{4}{3}\left(\frac{k}{k_{eq}}\right)+\frac{5}{2}\left(\frac{k}{k_{eq}}\right)^2\right]\mathcal{P}_{T}(k).
\ena
Among the terms in the square bracket, the last term dominates for the gravitational waves with $k\gg k_{eq}$  while the first one dominates for the gravitational waves with $k \ll k_{eq}$ which correspond to modes that reentered the horizon during the radiation- and the matter-dominated era, respectively.

The primordial tensor power spectrum can always be described as
\bea\label{tensorPS}
\mathcal{P}_{T}(k)=\mathcal{P}_{T}(k_{\ast})\left(\frac{k}{k_{\ast}}\right)^{n_T+\frac{\alpha_T}{2}\ln (k/k_{\ast})},
\ena
where $\alpha_T\equiv dn_{T}/d\ln k$ is the running of tensor spectral index $n_{T}$. By using the standard slow-roll analysis in the single-field inflation model, we obtain the following relations between observable quantities and slow-roll parameters:
\bea\label{obspar}
n_{T}\simeq-\frac{r}{8}, \quad r\simeq-\frac{8}{3}(n_S-1)+\frac{16}{3}\eta_V, \quad  \alpha_{T}\simeq\frac{r}{8}\left[(n_S-1)+\frac{r}{8}\right],
\ena
where $r\equiv \mathcal{P}_{T}(k_\ast)/\mathcal{P}_{S}(k_\ast)$ is the tensor-to-scalar ratio at the pivot scale $k_\ast$.
Although the tensor component has not yet been observed in CMB anisotropies, the amplitude for the scalar component has been determined quite accurately at a certain pivot scale. By substituting Eqs. (\ref{tensorPS})--(\ref{obspar}) into Eq. (\ref{h02OGW}), we obtain the energy spectrum of the primordial gravitational waves in terms of the physical frequency, $f=k/2\pi$, observed today
\bea\label{ESofGWofk}
h_0^2\O_{GW}=\frac{3h_0^2}{4H_0^2\tau_0^4} \left(\frac{\O_m}{\O_{\Lambda}}\right)^2 \frac{1}{f^2} \left[1+\frac{4}{3}\left(\frac{f}{f_{eq}}\right)+\frac{5}{2}\left(\frac{f}{f_{eq}}\right)^2\right] \mathcal{P}_{S}(k_\ast)r\left(\frac{f}{f_{\ast}}\right)^{n_T+\frac{\alpha_T}{2}\ln (k/k_{\ast})},
\ena
where we use $f_{eq}\equiv k_{eq}/2\pi$ and $f_{\ast}\equiv k_\ast/2\pi$. We will use this result for the single-field inflation model later in the next section to compare with a result for the space-condensate inflation model.

\section{Gravitational waves induced by the space-condensate inflation model}\label{sec3}

In this section, we consider the space-condensate inflation model \cite{PhysRevD.88.043523, underprep, PhysRevD.91.063521} as a source for the primordial gravitational waves. We consider the following action with the additional kinetic terms for the triad of scalar fields, $\sigma^a$, that are symmetric under $SO(3)$:
\bea\label{eq:action12}
S=\int d^4x\sqrt{-g}\left(\frac{M_p^2}{2}R-\frac{1}{2}g^{\mu\nu}\partial_{\mu}\phi\partial_{\nu}\phi -\frac{1}{2}g^{\mu\nu}\partial_{\mu}\sigma^a\partial_{\nu}\sigma^a-V(\phi)\right).
\ena
By varying the action, Eq. (\ref{eq:action12}), one can obtain the equations of motion of the scalar fields $\phi$ and $\sigma$ as
\bea\label{eq:beomscalar}
&&\partial_\mu\partial^{\mu}\phi-V_{\phi}=0,\\
&&\frac{1}{\sqrt{-g}}\partial_\mu\left(\sqrt{-g}g^{\mu\nu}\partial_\nu\sigma^a\right)=0,\label{eq:beomsigma}
\ena
where $V_{\phi}\equiv dV/d\phi$. To preserve the cosmological principles of homogeneity and isotropy for this model in the flat FLRW background with metric, $ds^2=-dt^2+a(t)^2(dx^2+dy^2+dz^2)$, we choose a spatially linear background solution for $\sigma^a$ \cite{ArmendarizPicon:2007nr, Endlich:2012pz, 'tHooft:2007bf, Lee:2009zv, PhysRevD.88.043523, underprep, PhysRevD.91.063521} ,
\bea
\sigma^a=\xi x^a,
\ena
where $\xi$ is an arbitrary constant of mass dimension two. By considering the flat FLRW metric, along with the spatially linear configuration of $\sigma$ fields that are symmetric under $SO(3)$, the spatial-condensate inflation model \cite{PhysRevD.88.043523, underprep, PhysRevD.91.063521} can be constructed. Thus, the background equations of motion are obtained as
\bea
H^2&=&\frac{1}{3M_p^2}\left(\frac{1}{2}\dot{\phi}^2+V+\frac{3\xi^2}{2a^2}\right),\label{huble00}\\
\dot{H}&=&-\frac{1}{2M_p^2}\left(\dot{\phi}^2+\frac{\xi^2}{a^2}\right),\label{hubleij}\\
0&=&\ddot{\phi}+3H\dot{\phi}+V_{\phi}.\label{fieldeq}
\ena

It can be easily seen in Eqs. (\ref{huble00})--(\ref{hubleij}) that the effect of the $\xi$-dependent term, decreases rapidly as the scale factor increases exponentially during inflation. If the $\xi$-dependent term in Eq. (\ref{huble00}) is much larger than the potential term at the onset of the slow-roll inflation, then the comoving horizon remains constant,
\bea
(aH)^{-1} \simeq \left(\frac{\xi^2}{2M_p^2}\right)^{-\frac{1}{2}},
\ena
until the potential term becomes comparable to the $\xi$-dependent term. Therefore, there exists the early phase transition during inflation at which the potential term starts dominating over the $\xi$-dependent term such that usual slow-roll inflation takes place. At the end of the inflation, this model asymptotically approaches the standard single-field inflation model with the small contribution from the $\xi$-dependent term \cite{PhysRevD.88.043523, underprep, PhysRevD.91.063521}.

In the context of the slow-roll inflation, a system is governed by the slow-roll parameters such that the slow-roll approximations can be reflected. In our model, following Refs.~\cite{PhysRevD.88.043523, underprep}, we obtain the slow-roll parameters
\bea\label{slprms}
\epsilon_{V}\simeq\frac{M_p^2}{2}\left(\frac{V_{\phi}}{V}\right)^2\left(1-\frac{3\xi^2M_p^2}{a^2V}\right), \quad \eta_{V}\simeq\frac{V_{\phi\phi}}{V}M_p^2\left(1-\frac{3\xi^2M_p^2}{2a^2V}\right),
\ena
and we will show the observable quantities in terms of these slow-roll parameters. Before moving on further, it is worth pointing out the main difference of the space-condensate inflation model compared to the single-field inflation model at the linear perturbation for the tensor mode.

In the linearly perturbed background with metric (\ref{FRWmetric}), the linearized evolution equation of the primordial gravitational waves can be obtained for the space-condensate inflation model as
\bea\label{eq:tenspert}
h_{\l,k}''+2\mathcal{H}h_{\l,k}'+\left(k^2+\frac{2\xi^2}{M_p^2}\right)h_{\l,k}=0,
\ena
where we have an additional term, $2\xi^2/M_p^2$, in the bracket, which makes our model different from the single-field inflation model described in Eq. (\ref{waveeq1}). The power spectrum of tensor modes is obtained as
\bea
\mathcal{P}_{T}(k)\simeq&\frac{16 H_{\ast}^2}{\pi M_p^2}\left(1+(2C-2)\epsilon-\frac{37\tilde{\xi}^2}{12 M_p^2 k^2}\right)\label{powert},
\ena
where ``$\,^\ast$" indicates the value at the horizon crossing point, $k=aH$. The observable quantities for the space-condensate inflation model, therefore, were obtained in Ref. \cite{underprep} as follows:
\bea
&n_S-1\simeq 2\eta_V-6\epsilon_V+\frac{2\xi^2}{\epsilon_V k^2M_p^2}, \quad n_T\simeq -2\epsilon_V+\frac{31\xi^2}{6 k^2M_p^2}, \quad r\simeq16\epsilon_V+\frac{16\xi^2}{k^2M_p^2},\\
&\alpha_S=-\frac{4\xi^2}{\epsilon_V k^2M_p^2}, \quad \alpha_T=-\frac{31\xi^2}{3k^2M_p^2}.
\ena

Since our model asymptotically approaches the single-field inflationary model at the end of inflation, we can treat our model as a single-field inflation model with a small contribution from the $\xi$-dependent term.
Therefore, by substituting these observable quantities into Eq. (\ref{ESofGWofk}), we obtain the energy spectrum of the primordial gravitational waves in terms of the physical frequency observed today
\bea\label{Omegafwhole}
h_0^2\O_{GW}=\frac{3h_0^2}{16 \pi^2 H_0^2\tau_0^4} \left(\frac{\O_m}{\O_{\Lambda}}\right)^2 \frac{1}{f^2} \left[1+\frac{4}{3}\left(\frac{f}{f_{eq}}\right)+\frac{5}{2}\left(\frac{f}{f_{eq}}\right)^2\right]\nn
\times\mathcal{P}_{S}(k_\ast)r\left(\frac{f}{f_{\ast}}\right)^{-\frac{r}{8}+\frac{43\xi^2}{24\pi^2M_p^2f^2}-\frac{31\xi^2}{24\pi^2M_p^2f^2}\ln (f/f_{\ast})},
\ena
where $f_{eq}\equiv k_{eq}/2\pi$ and $f_{\ast}\equiv k_\ast/2\pi$.

For the numerical result below, we use the standard model of cosmology and the cosmological parameters adopted in this paper are listed in Table \ref{table:1}  \cite{Ade:2015xua, Ade:2015lrj}.

\begin{figure}[H]
\centering
\includegraphics[width=.8\textwidth]{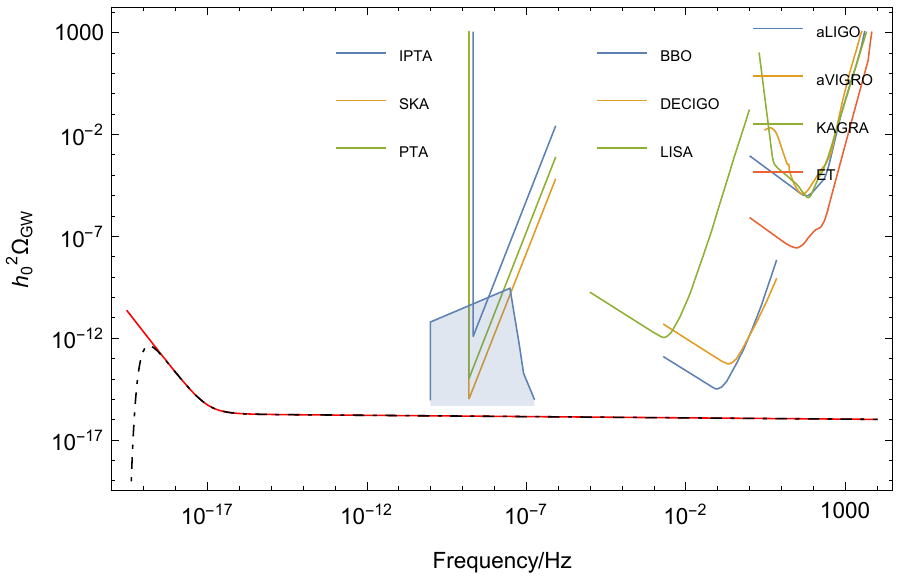}
\caption{The energy spectrum of primordial gravitational waves versus the gravitational-wave frequency. The solid line is for Eq. (\ref{ESofGWofk}) while the dot-dashed one is for Eq. (\ref{Omegafwhole}). We choose $r_{0.002}=10^{-1}$ for both the solid and the dot-dashed line with $\xi=0$ and $\xi=10^{-19}$ respectively. The observational upper bounds of the gravitational-wave background from several different observations are provided as well.}
\label{fig:fig1}
\end{figure}

\begin{figure}[H]
\centering
\subfigure[]
{\includegraphics[scale=0.6]{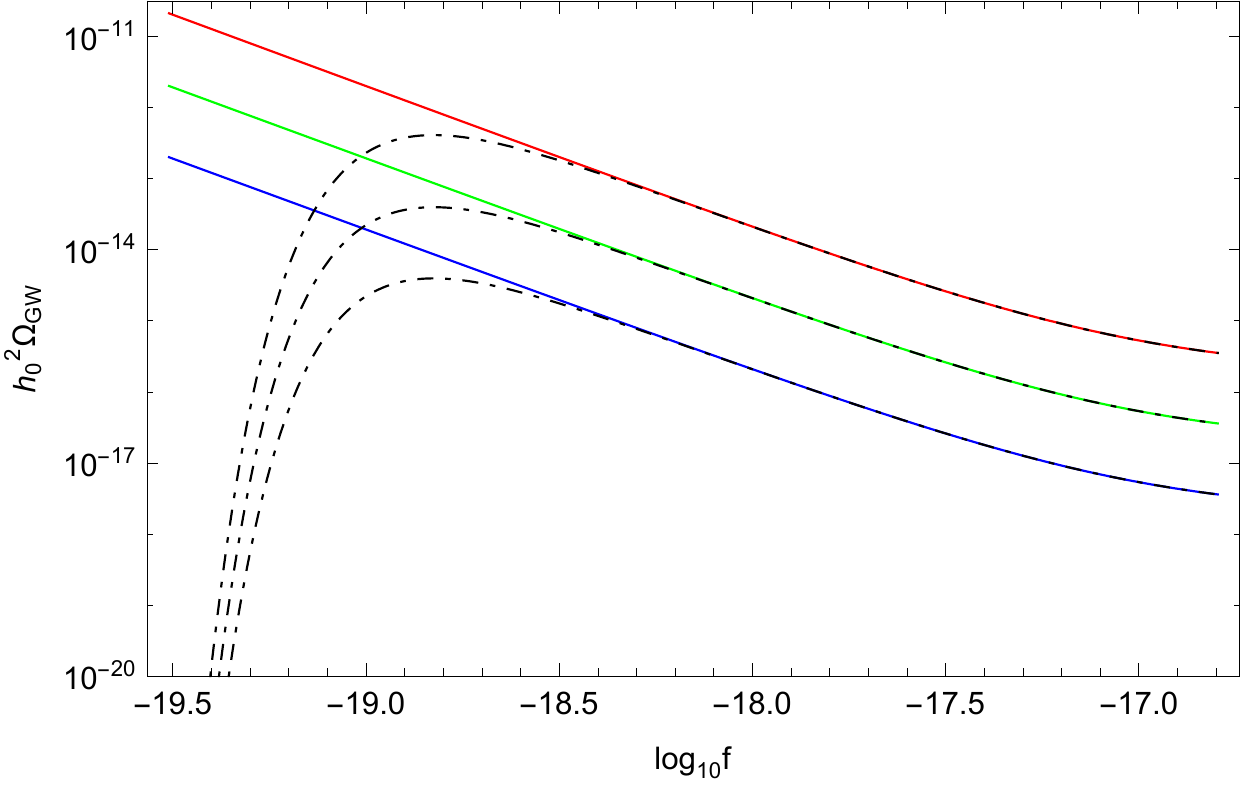}\label{fig:fig2c}}
\subfigure[]
{\includegraphics[scale=0.6]{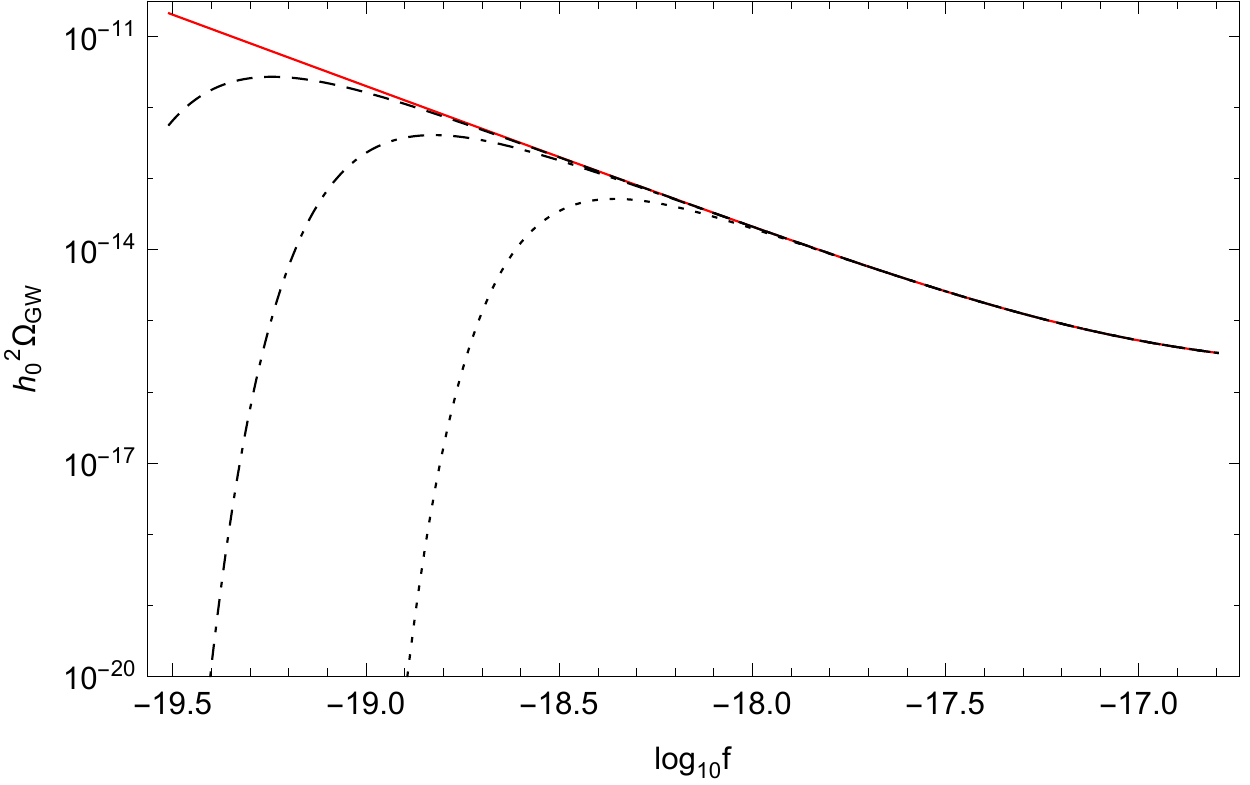}\label{fig:fig2d}}
\caption{The low-frequency range, $f<f_{eq}$, of Fig.~\ref{fig:fig1}. (a): The model parameter is fixed to $\xi=10^{-19}$ and three different values of the tensor-to-scalar ratio, including $r=10^{-1}$ (red), $10^{-2}$ (green) and $10^{-3}$ (blue), are considered. (b): The tensor-to-scalar ratio is fixed to $r=10^{-1}$ and three different values of the model parameter, including $\xi=3.1\times10^{-20}$ (dashed), $10^{-19}$ (dot-dashed) and $4\times10^{-19}$ (dotted), are considered.}\label{fig:fig2}
\end{figure}

\begin{table}[h!]
\caption{The list of parameters.}
\centering
\begin{tabular}{||c | c | c||}
 \hline
 $H_0=100 h_0\,\text{km}\,\text{s}^{-1}\,\text{Mpc}^{-1}$  &  $k_{eq}=0.073 \Omega_m\, h_0^2\,\text{Mpc}^{-1}$ &  $\mathcal{P}_{S}(k_{\ast})\simeq 2.19\times10^{-9}$ \\
 $h_0\simeq0.6731$ & $\Omega_{m}=0.315$ &  $k_\ast=0.002\text{Mpc}^{-1}$ \\
  $\tau_0=1.41\times10^4$Mpc   & $\Omega_{\Lambda}=0.685$ & $f_{\ast}=3.092\times10^{-18}\text{Hz}$ \\[1ex]
 \hline
\end{tabular}
\label{table:1}
\end{table}
Figure  \ref{fig:fig1} shows the energy spectrum of primordial gravitational waves, obtained by both Eq. (\ref{ESofGWofk}) and Eq. (\ref{Omegafwhole}), as a function of frequency $f$. We choose $r_{0.002}=10^{-1}$ for both the solid line and the dot-dashed line since recent Planck 2015 data favor $r_{0.002}<10^{-1}$ for the upper bound on the tensor-to-scalar ratio \cite{Ade:2015lrj}. The solid line represents the case where $\xi=0$ while the dot-dashed line represents the case with $\xi=10^{-19}$. We add the sensitivity curves of several observations including the pulsar timing observation, and the space- and the ground-based laser interferometers to Fig.~\ref{fig:fig1} as a reference for the energy spectrum of gravitational waves. The sensitivity curves shown here are same as those provided in Ref.~\cite{0264-9381-32-1-015014}.
As seen in Fig.~\ref{fig:fig1}, the amplitude of the energy spectrum is still far below the observational upper bounds and the energy spectrum has a significant suppression in the low-frequency range. Before we realize the suppression of the energy spectrum, we would like to say a few words about the effects of both the tensor-to-scalar ratio $r$ and the model parameter $\xi$ on the energy spectrum. These effects are shown in Fig.~\ref{fig:fig2}. Figure \ref{fig:fig2c} shows that the change of the tensor-to-scalar ratio does not affect the suppression of the energy spectrum; instead it significantly affects the amplitude of energy spectrum.
On the other hand Fig.~\ref{fig:fig2d} shows that the change of the $\xi$ parameter plays an important role for the suppression of the energy spectrum. The suppression of the energy spectrum shifts to larger frequencies as the value of $\xi$ increases. As the value of $\xi$ decreases, the suppression shifts to smaller frequencies such that the energy spectrum of gravitational waves converges to that of the standard single-field inflation model. Therefore, it is clear that the suppression of the energy spectrum for the space-condensate inflation model depends on the value of the model parameter $\xi$.

In order to realize the suppression of the energy spectrum of the primordial gravitational waves induced by the space-condensate inflation model, we need to understand the early phase transition that takes a place during inflation. As we briefly mentioned earlier there exists an early phase before the usual slow-role inflation over which the $\xi$-dependent term dominates the potential term such that the comoving horizon remains constant $(aH)^{-1}\simeq (\xi/\sqrt{2}M_p)^{-1}$. The existence of a constant comoving horizon means that there also exists an observational lower bound \cite{PhysRevD.91.063521} for the comoving wave number, $k_{\text{min}}\simeq\xi/\sqrt{2}M_p$. Therefore, the gravitational-wave modes that satisfy $k\geq\xi/\sqrt{2}M_p$ can only cross the comoving horizon during the early phase transition, and the modes that satisfy $k<\xi/\sqrt{2}M_p$ never cross the comoving horizon and stay outside the horizon.

\begin{figure}[H]
\centering
\subfigure[]
{\includegraphics[scale=0.384]{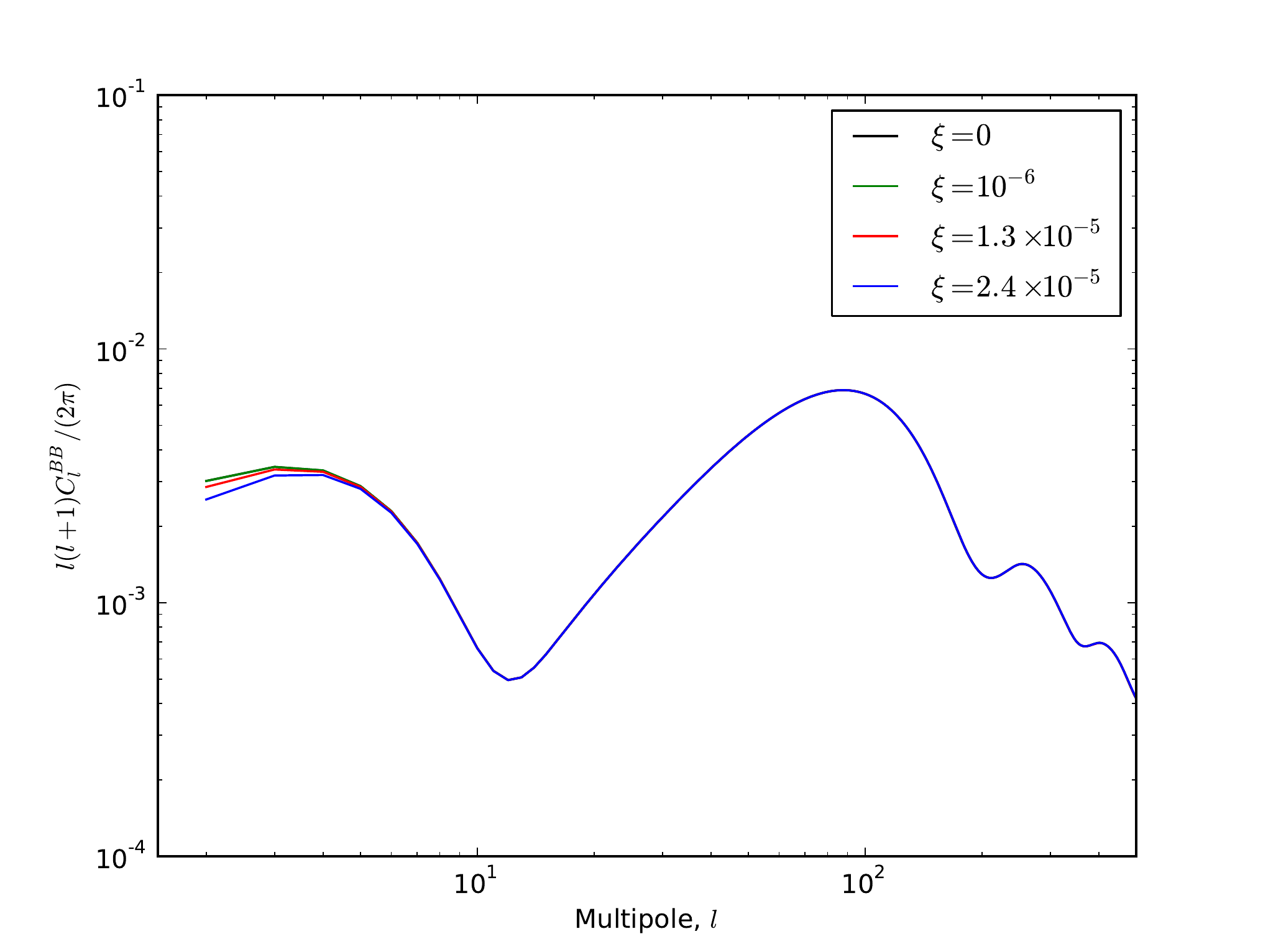}\label{fig:fig3a}}
\subfigure[]
{\includegraphics[scale=0.384]{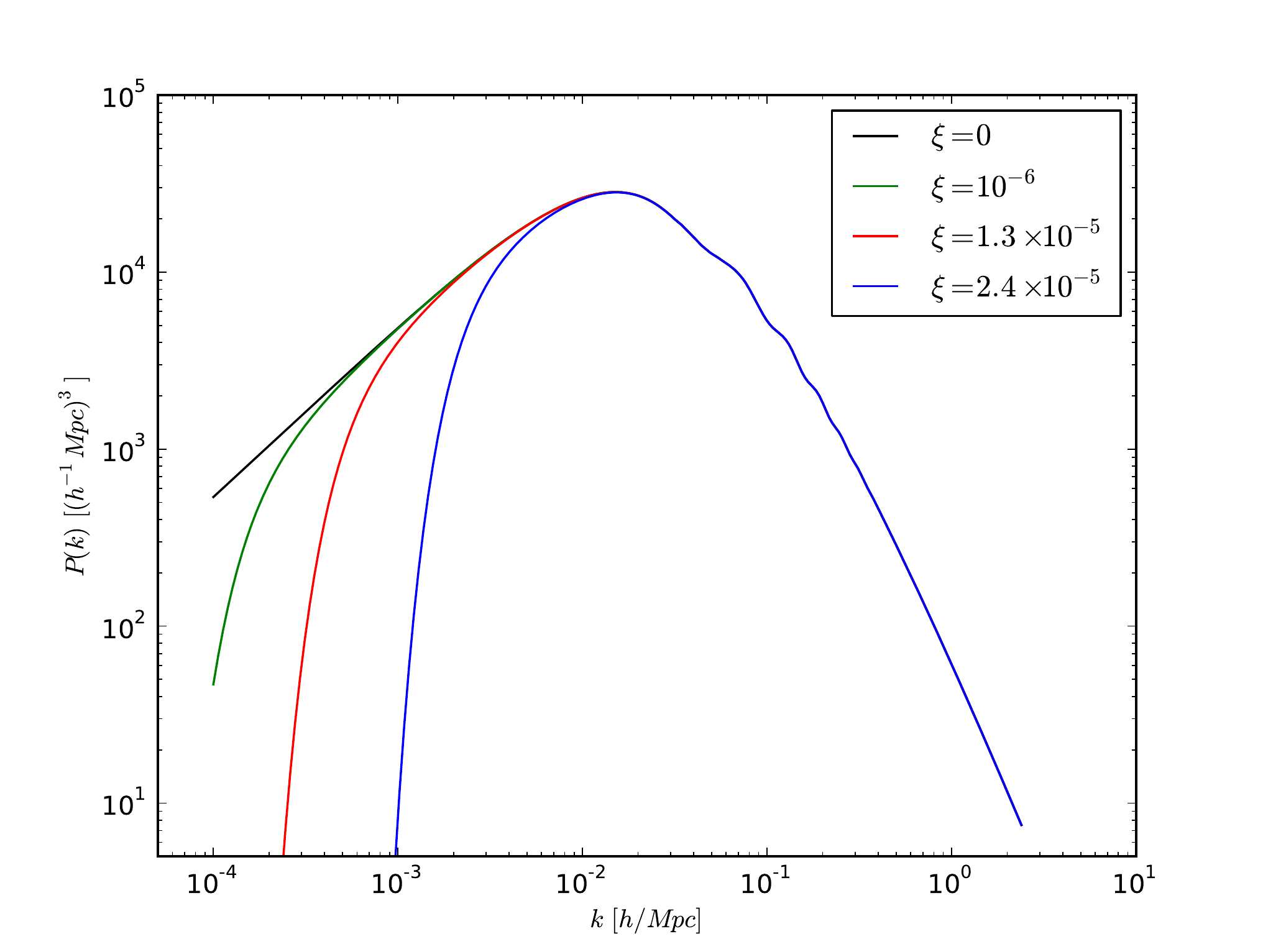}\label{fig:fig3b}}
\caption{The CMB angular power spectra (without lensing) for the B-mode polarization in (a) and the linear matter power spectra in (b), respectively. The spectra is calculated with the \texttt{CAMB} code \cite{CAMB} and we use the parameters given in Table \ref{table:1}.}\label{fig:fig3}
\end{figure}

If there exits an early phase just before the usual slow-roll inflation, say, that leads to the suppression of the energy spectrum of the primordial gravitational waves induced by the space-condensate inflation model at a certain frequency range depending on the value of $\xi$, then such a suppression can also be imprinted on the CMB measurement. This expectation is proven in Fig.~\ref{fig:fig3} where we plot both the CMB angular power spectra for the B-mode polarization and the linear matter power spectra for various different $\xi$ values. As seen in Fig.~\ref{fig:fig3}, the suppression happens as the $\xi$ parameter increases, or it converges to the standard result where $\xi=0$ as the $\xi$ parameter decreases.

\section{Discussion and conclusion}\label{sec4}

In this work, we have studied the primordial gravitational waves from the space-condensate inflation model. The gravitational-wave energy spectrum for our model was analytically calculated in Eq. (\ref{Omegafwhole}) as a function of frequency, while a similar result for the standard single-field inflation model was obtained in Eq. (\ref{ESofGWofk}). To obtain these analytic expressions, for simplicity, we have assumed that the phase transition between two consecutive regimes is abrupt such that only the cosmic expansion is responsible for the damping effect of the modes that reentered the Hubble horizon. The numerical result of our work has been shown in Fig.~\ref{fig:fig1} where we plot Eqs. (\ref{ESofGWofk}) and (\ref{Omegafwhole}) in light of several observational upper bounds, including the space- and ground-based laser interferometric detectors. As shown in Fig.~\ref{fig:fig1}, the energy spectrum of the primordial gravitational waves induced by the space-condensate inflation model is significantly suppressed as the frequency of the gravitational waves decreases, while it approaches that of the single-field inflation model as the frequency increases. In principle, the suppression of the energy spectrum occurs in any frequency range depending on the model parameter $\xi$. Fig.~\ref{fig:fig1} also shows that the amplitude of the energy spectrum is still far below the limits of current observational upper bounds in the detectable frequency range. In this work, the tensor-to-scalar ratio considered to have values $r_{0.002}\leq10^{-1}$ which is consistent with the Planck 2015 data \cite{Ade:2015lrj} .

Fig.~\ref{fig:fig2} shows the impact of both the tensor-to-scalar ratio $r$ and the model parameter $\xi$ on the energy spectrum of the primordial gravitational waves induced by the space-condensate inflation model. As shown in Fig.~\ref{fig:fig2c}, the change of the value of $r$ does not affect the suppression of the energy spectrum; instead it affects the amplitude of the energy spectrum. On the other hand, the parameter $\xi$ plays an important role for the suppression of the energy spectrum as shown in Fig.~\ref{fig:fig2d}. As the value of the parameter $\xi$ increases, the suppression of the energy spectrum shifts to a larger frequency range, or vice versa. The energy spectrum of primordial gravitational waves induced by the space-condensate inflation model converges to that of the single-filed inflation model as the $\xi$ parameter decreases.

In order to realize the suppression of the energy spectrum of the primordial gravitational waves for the space-condensate inflation model, we have considered its features by using the background equations of motion, Eqs. (\ref{huble00})--(\ref{fieldeq}). For this model, there exits an early phase transition during inflation from a phase of background evolution over which the $\xi$-dependent term dominates the potential term to a phase of the usual slow-roll inflation where the potential term dominates over the $\xi$-dependent term. The existence of the phase transition during the inflationary epoch leads to the existence of the observational lower bound for the comoving wave number, $k_{\text{min}}\sim \xi/\sqrt{2}M_p$, and that observational lower bound is then responsible for the suppression of the energy spectrum of the primordial gravitational waves for the space-condensate inflation model. The modes with wave number smaller than the observational lower bound, $k\ll k_{\text{min}}$, would never cross the comoving horizon, and they stay outside the horizon. Those modes are causally disconnected from our Universe; that is why we have seen the suppression effect in the energy spectrum of the primordial gravitational waves. Therefore, the suppression of the energy spectrum takes place when the gravitational wave frequency gets smaller than a certain value, $f\lesssim f_{\text{min}}$ where $f_{\text{min}}\simeq \xi/\sqrt{2}M_p$. In other words, no suppression occurring as long as the frequency is larger than $f_{\text{min}}$ such that $f\gg f_{\text{min}}$.

In this work we have considered the cosmic expansion as the only damping effect for the modes reentering the Hubble horizon. Therefore, it would be interesting to take other damping effects into account.

\section*{Acknowledgements}
We appreciate APCTP for its hospitality during completion of this work. We acknowledge the use of the publicly available {\tt CAMB-code}.\, G.T. thanks Sunly Khimphun for useful help.\, S.K. was supported by the Basic Science Research Program through the NRF funded by the Ministry of Education (No. NRF-2014R1A1A2059080). B.H.L. was supported by the National Research Foundation of Korea (NRF) grant funded by
the Korean government(MSIP) No. 2014R1A2A1A01002306 (ERND).

\bibliography{ref}{}

\bibliographystyle{ieeetr}

\end{document}